# Titan Atmospheric Chemistry Revealed by Low-temperature $N_2$-$CH_4$ Plasma Discharge Experiments

Chao He[1*], Joseph Serigano[1], Sarah M. Hörst[1,2], Michael Radke[1], Joshua A. Sebree[3]

[1]Department of Earth and Planetary Sciences, Johns Hopkins University, Baltimore, MD, 21218, USA che13@jhu.edu

[2]Space Telescope Science Institute, Baltimore, MD, 21218, USA

[3]Department of Chemistry and Biochemistry, University of Northern Iowa, Cedar Falls, IA, 50614, USA

**Abstract**

Chemistry in Titan's $N_2$-$CH_4$ atmosphere produces complex organic aerosols. The chemical processes and the resulting organic compounds are still far from understood, although extensive observations, laboratory, and theoretical simulations have greatly improved physical and chemical constraints on Titan's atmosphere. Here, we conduct a series of Titan atmosphere simulation experiments with $N_2$-$CH_4$ gas mixtures and investigate the effect of initial $CH_4$ ratio, pressure, and flow rate on the production rates and composition of the gas and solid products at a Titan relevant temperature (100 K) for the first time. We find that the production rate of the gas and solid products increases with increasing $CH_4$ ratio. The nitrogen-containing species have much higher yield than hydrocarbons in the gas products, and the N/C ratio of the solid products appears to be the highest compared to previous plasma simulations with the same $CH_4$ ratio. The greater degree of nitrogen incorporation in the low temperature simulation experiments suggests temperature may play an important role in nitrogen incorporation in Titan's cold atmosphere. We also find that $H_2$ is the dominant gas product and serves as an indicator of the production rate of new organic molecules in the experiment, and that $CH_2NH$ may greatly contribute to the incorporation of both carbon and nitrogen into the solid particles. The pressure and flow rate affect the amount of time of the gas mixture exposed to the energy source and therefore impact the $N_2$-$CH_4$ chemistry initiated by the plasma discharge, emphasizing the influence of the energy flux in Titan atmospheric chemistry.

**Keywords:** Titan tholins, Organic aerosols, Photochemistry, Prebiotic chemistry Nitrogen incorporation, Laboratory simulations

1. INTRODUCTION

The thick atmosphere of Titan is primarily made up of nitrogen and methane. Complex organic chemistry happening in this $N_2$-$CH_4$ atmosphere produces a large number of organic molecules, leading to the formation of Titan's characteristic organic aerosols.[1-5] The complex organic chemistry is of great interest because it may result in the formation of prebiotic molecules required for the origin of life.[6-7] Further chemical evolution may happen when the organic molecules interact with surface periodic liquid water.[8] Therefore, studying the organic chemistry happening on Titan would help to understand prebiotic chemistry and perhaps life's origin on Earth.

Although the successful Cassini-Huygens mission has improved our understanding of many processes on Titan, the organic chemistry in Titan's atmosphere is not well understood, especially the detailed chemical processes from gas molecules ($N_2$ and $CH_4$) to aerosol particles with poorly known composition. Both numerical models and laboratory experiments have been employed to simulate and study these complex chemical processes. However, numerical models usually only consider the gas phase chemistry up to C6 species[9] and are subject to larger uncertainties due to lack of necessary chemical kinetic data. Laboratory experiments can help to constrain the chemical pathways that produce more complex aerosols. For over four decades, different experimental setups have been developed to simulate Titan's atmospheric chemistry by exposing $N_2$-$CH_4$ gas mixtures to a variety of energy sources under different conditions. The resulting solid particles, so called "tholins", serve as analogues of Titan's aerosols. The compositions of both gas and the solid products are usually characterized to study the chemical processes in Titan's atmosphere. A detailed discussion of previous Titan simulation experiments, including different setups, various energy sources, and properties of tholins, can be found in two comprehensive reviews.[3-4] The measurements by the Cassini Ion and Neutral Mass Spectrometer (INMS) and the Aerosol Collector Pyrolyzer (ACP) on Huygens have confirmed that nitrogen incorporates into both gas and aerosols to a significant extent in Titan's atmosphere.[10-12] Plasma discharge is energetic enough to directly dissociate stable $N_2$ molecules and therefore is widely used in such simulation experiments to study the nitrogen incorporation.[3,4]

Because each individual laboratory uses different experimental setup and analysis techniques in their simulation experiments, it is difficult to directly compare the results across laboratories. However, performing simulation experiments on the same setup with varied parameters (gas mixture, pressure, and flow rate) allows investigation of the effects of different experimental conditions on the compositions and properties of the resulting gas and solid products. For example, several groups have studied how $CH_4$ ratio in the initial gas mixture affects the degree of nitrogen incorporation or the gas to solid particle conversion efficiency.[13-16] Pressure affects the collision frequencies of reactive species and therefore impacts the atmospheric chemistry. For instance, Alcouffe et al.[17] have shown that solid particles are formed faster at higher pressure over the range of 0.2 to 3 mbar, while Imanaka et al.[18] have demonstrated that nitrogen is more likely to incorporate into carbon networks in tholins formed at lower pressures.

The Planetary HAZE Research (PHAZER) chamber at Johns Hopkins University can operate at various conditions and has been used to simulate a range of planetary atmospheres.[19-25] The PHAZER chamber is one of few setups[19,26,27] that can operate at cryogenic temperature to simulate Titan's cold atmosphere environment. The other low temperature setup that can simulate Titan's low temperature atmospheric chemistry is the COSmIC/THS chamber at NASA Ames.[27] The cooling mechanism of these two chambers is different for running low temperature experiments. The gas is cooled by flowing through a cooling coil in liquid nitrogen with the PHAZER chamber;[19] while the gas is cooled through free-jet supersonic expansion with the COSmIC/THS chamber.[27] Temperature is a crucial factor for atmospheric chemistry simulations because it not only determines the species available for reactions by impacting the stability and condensation of gas molecules, but also affects reaction rate constants and heterogeneous adsorption of gases onto particle surfaces.[4] In this study, we conduct a series of Titan atmosphere simulation experiments by exposing a range of $N_2$-$CH_4$ gas mixtures to plasma discharge at low temperature with the PHAZER chamber. Plasma discharges and UV photons are the two main energy sources used for initiating the chemistry in Titan simulation experiments.[3,4] Solar UV flux is the main driver for Titan's atmospheric chemistry, but charged particles (i.e., cosmic rays, magnetospheric protons, and electrons) also contribute to the energy inputs greatly, especially when Titan moves across the Saturnian magnetopause and is fully exposed to solar wind.[4] Plasma discharges may be able to simulate the chemistry induced by the charged particles. Moreover, the energy distribution of the plasma discharges provides a closer simulation of the solar spectrum energy distribution than that offered by UV photon sources.[4,27] Laboratory experiments with UV photons usually employ a typical Lyman alpha emission lamp[4,7,16] or EUV photons at selective wavelengths (60/82.5 nm[28], 73.6 nm[29], 85.6 nm[30], and 89.2 nm[31]). The plasma discharges are not only energetic enough to dissociate $N_2$ directly (while a typical Lyman alpha emission lamp could not), but also offer a wider energy distribution than EUV sources. Another benefit of using plasma discharges is that the plasma methods usually produce larger amounts of aerosol analogues than the UV methods, allowing analyses that require more sample. The mixing ratio of $CH_4$ in Titan's atmosphere ranges from 5.6% at the surface to 1.5% in the upper atmosphere.[5] We run our experiments with different mixing $CH_4$ ratios (0.5, 1, 2, 5, and 10%) to study how it will affect the chemistry, which may also provide insights into the chemical processes happening at different altitudes in Titan's atmosphere. We determine the gas composition through deconvolution of the mass spectra measured with an in situ mass spectrometer and quantify the production rate and elemental composition of the solid products. These results allow us to study the influence of $CH_4$ concentration, pressure, and flow rate on the production rates and composition of the gas and solid products at low temperature for the first time.

## 2. MATERIALS AND METHODS

*2.1. Experimental setup and procedure*

A schematic of the PHAZER chamber is shown in Figure 1 and a detailed description of the setup can be found in our previous study.[19] The experimental procedure in this study is briefly described below. Different concentrations (0.5, 1, 2, 5, and 10%) of $CH_4$ (purity: 99.999%, Airgas) in $N_2$ (purity: 99.999%, Airgas) gas mixtures are premixed in a stainless-steel cylinder. The gas mixture is cooled by flowing through a cooling coil that is immersed in liquid nitrogen (77 K). After the cooling coil, the temperature of the gas mixture is determined to be ~100 K based on the ideal gas law ($P_1/P_2=T_1/T_2$). The cold gas mixture is then exposed to AC glow discharge that is a cold plasma source. In such discharge, only a small fraction of the gas gets ionized, which does not significantly increase the total gas temperature in the system.[4,27] The gas mixture continually flows through the reaction chamber and then is pumped out by a dry scroll pump (Agilent, IDP-3). We monitor the composition of the gas mixture flowing out of the chamber using a Residual Gas Analyzer (RGA, Stanford Research Systems, RGA300, a quadrupole mass spectrometer). The procedure for the mass spectral measurements and data processes is described in Section 2.4.

We run the experiments for 72 hours, and then wait another 48 hours to allow the chamber to warm up and the volatile components to be pumped out before collecting solid samples in a dry, oxygen-free $N_2$ glove box (InertCorp.). We weigh the collected solids using an analytical balance (Sartorius Entris 224-1S with standard deviation of 0.1 mg), and keep them in the glovebox before further analysis. Before each experiment, the chamber is cleaned using detergent, solvents, and ultrasonic cleaner, and then is baked at 110 °C overnight. The chamber is pumped down to $1\times10^{-4}$ torr before starting an experiment, and the base pressure of the RGA chamber is $5\times10^{-7}$ torr.

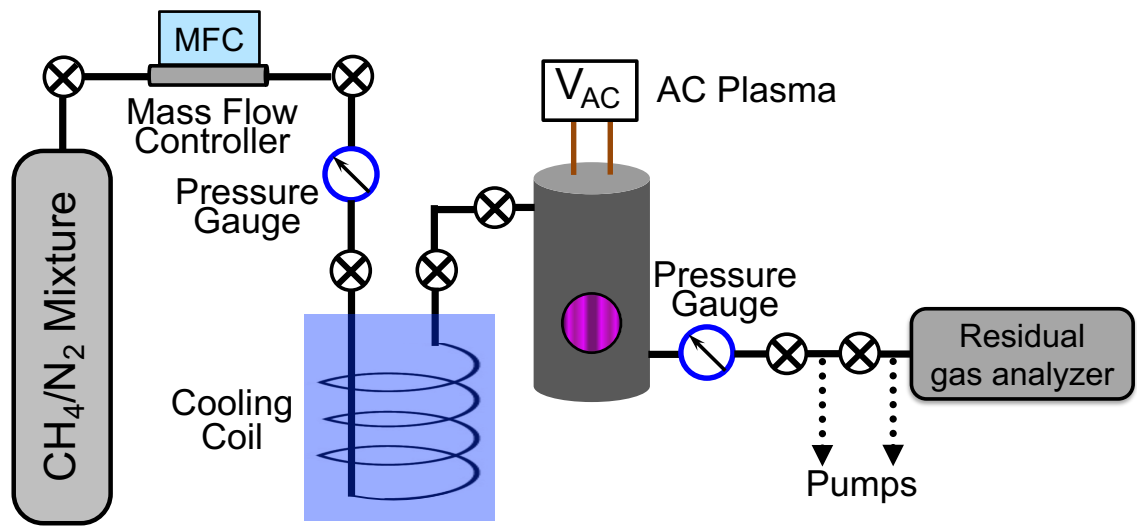

Figure 1. Schematic of the experimental setup (Modified from He et al. 2017)[19].

*2.2. Experimental conditions*

Table 1. Experimental conditions

| Temperature (K) | Flow rate (sccm) | Pressure (Torr) | Mixing ratio of $CH_4$ (%) | Running time (hour) |
|---|---|---|---|---|
| 100 | 10 | 2 | 0.5, 1, 2, 5, 10 | 72 |
| 100 | 2.5 | 1, 3, 5 | 10 | 4 |
| 100 | 5 | 1.5, 3, 5 | 10 | 4 |
| 100 | 10 | 2, 3, 5 | 10 | 4 |

The experimental conditions are listed in Table 1. For all our experiments, the temperature of the gas mixture is ~100 K to mimic the low temperature condition of Titan's atmosphere. The flow rate of the gas mixture varies from 2.5 to 10 sccm. At each given flow rate (2.5, 5, or 10 sccm), we adjust the pump speed to change the pressure in the chamber (1-5 Torr). Titan's atmospheric chemistry starts from >1000 km altitude (140-150 K and ~$10^{-7}$-$10^{-8}$ torr) and continues throughout the whole atmosphere (94 K and 1.5 bar at the surface).[5] The pressure (1 to 5 torr) in our experiments is close to the pressure in Titan's stratosphere (100-210 km).[5] In practice, laboratory simulation usually requires pressure higher than Titan's upper atmosphere to enhance the chemistry, in order to yield sufficient product for analysis in a reasonable time scale. Titan simulation experiments with plasma discharge are usually performed at pressure similar to that we use here or at higher pressure.[1-4,13-19] In addition, our previous study has shown that the optical properties of the aerosol analogues produced in our experiments are close to the retrieved extinction coefficients (k) of Titan's aerosols in the near IR region observed by the Cassini–Huygens mission.[32] For the experiments with different $CH_4$ ratios, we fix the flow rate (10 sccm) and pressure (2 Torr), and run the experiments for 72 hours to collect solid samples. For other experiments, we run the experiments for 3 hours and only monitor the gas products with RGA.

*2.3. Elemental analysis of the solids*

For those experiments that we collect solid samples, we determine the elemental compositions of each sample by using an organic elemental analyzer (Flash 2000 Elemental Analyzer, Thermo Scientific). We first weigh a small amount (~1 to 2 mg) of solid sample in a tin capsule, and then place the capsule enclosing the sample in the analyzer for elemental analysis. We repeat the measurement 3 times for each sample. The contents of carbon, hydrogen, and nitrogen elements are determined directly from the measurement, and the percentage of O is calculated by mass subtraction. Since the initial gas mixture only contains $CH_4$ and $N_2$, the resulting solid material should only contain C, H, and N atoms. Samples are kept under dry $N_2$ atmosphere until weighing for elemental analysis; however, oxygen could come from the adsorbed water vapor during the short weighing period (a few minutes).

*2.4. Mass spectral measurements of the gas products and data process procedure*

A small portion of the gas mixture flowing out of the chamber is channeled into the RGA and the mass spectrum (MS) of the gas mixture is recorded. The RGA is linked to the chamber through 30 cm stainless-steel tube. A stainless-steel needle valve is used to control the amount of gas flowing into the RGA chamber. The aperture of our needle valve (1.4 mm) is about two orders of magnitudes larger than the mean free path of the gas in our experiments (0.004-0.02 mm), therefore mass fractionation due to effusion can be neglected. We use the electron ionization (EI) source with a standard 70 eV energy for the MS measurements. In each experiment, we first record the mass spectrum (50-scan average) of the original gas mixture when the plasma is off; after turning on the plasma, we wait 30 minutes and then collect the mass spectrum (100-scan average) of the gas mixture that includes the unreacted $CH_4/N_2$ and newly formed gas products. We then normalize these mass spectra (MS of the original gas mixture and MS of the gas mixture when the plasma is on) relative to the intensity of 28 amu in order to compare the spectra and determine peaks with significant intensity changes. The peak at 28 amu mainly comes from $N_2$ that accounts for at least 90% in the initial gas mixture, and its peak intensity is constant enough to be used as a fixed reference.[19,33]

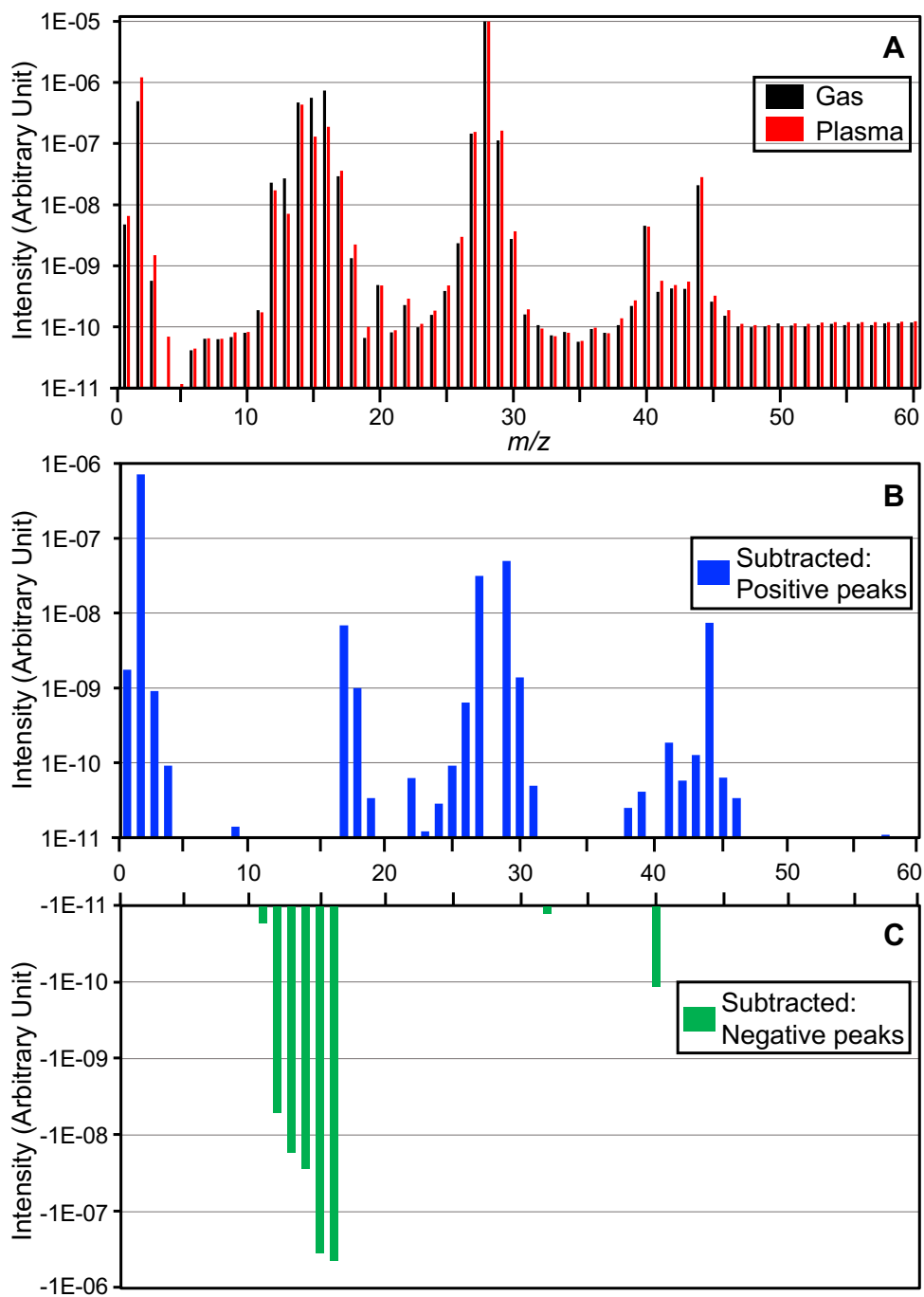

Figure 2. (A) The mass spectra of the original gas mixture (black) and the gas mixture with the plasma on (red) in the 10%-CH$_4$-10-sccm-2-Torr experiment. Peaks from 1 to 60 amu are shown since the heavier peaks are near the noise level; (B) the positive peaks and (C) the negative peaks of the subtracted mass spectrum from the same experiment.

Figure 2A shows the normalized MS of the original gas mixture (black) and the gas mixture with the plasma on (red) in our 10%-CH$_4$-10-sccm-2-Torr experiment. In Figure 2A, there are a few peaks corresponding to air components, such as 18 m/z ($H_2O$), 40 m/z (Ar), and 44 m/z ($CO_2$). We quantify the mixing ratio of $CH_4$, $N_2$, $H_2O$, $O_2$, Ar, and $CO_2$ in the initial gas mixture using the Monte Carlo approach described in Section 2.5. Some of these species could be due to micro-leaks and/or residual air components in the RGA chamber and reactor. $CO_2$ in our spectra is mainly due to residual in the RGA chamber. Major oxygen species in Earth's atmosphere, $O_2$ and $H_2O$, might potentially introduce oxygen to the chemical system.[31] The residual $O_2$ and $H_2O$ could potentially impact to the chemistry in our flowing system. However, the mixing ratios of $O_2$ and $H_2O$ (at $1\times10^{-5}$ or $2\times10^{-4}$ level) are much lower than the main gas mixture ($CH_4$ and $N_2$); therefore, the oxygenated species are not expected as major products in our experiments. After normalization, we obtain the subtracted mass spectrum for each experiment by subtracting the MS of the original gas mixture from the MS of the gas mixture with the plasma on. In a subtracted mass spectrum, the negative peaks are caused by the destruction of the initial gas species, while the positive peaks are due to the formation of new species. In our subtracted mass spectra (Figure 2C), the main negative peaks are at 12-16 amu, due to the depleting of $CH_4$ in the system. Figure 2B shows the subtracted mass spectrum from the 10%-$CH_4$-10-sccm-2-Torr experiments. The unit resolution of the RGA does not allow us to resolve different species by their exact mass because the mass signals of different species overlap with each other. To identify and quantify the newly formed species in a mixture, we need to deconvolve the mass spectra.

*2.5 Deconvolution of Mass Spectrum*

Ideally, we can determine the mixing ratios of each species using the calibration fragmentation patterns on the RGA to solve a linear combination of each species' concentration multiplied by their fragmentation patterns. However, such calibration data are not available. Instead, we use calibration data from the National Institute of Standard and Technology (NIST) mass spectral library.[34] Because the fragmentation patterns vary among different instruments, we use a Monte Carlo based approach to vary the peak intensity of individual fragment ions for each species. This approach is similar to that developed by in Gautier et al.[35], which has been used for analyzing both lab RGA data[31] and INMS data from the Cassino mission.[36,37] In this study, we update the fitting procedure to use the MultiNest nested sampling algorithm[38], which significantly reduces computational time. Instead of drawing new samples from a set of bounding ellipsoids, which is the default MultiNest sampling procedure, we use the MultiNest MCMC sampling procedure based on that of Veitch and Vecchio.[39] During MCMC sampling, we allow fragmentation peak intensities to vary by ±30% following previous studies.[31,35-37] Although we allow for ±30% fragment intensity variations, the final results for the most significant fragments vary only by a few percent from the original reference fragment intensity from NIST calibration data. When running the algorithm we use 100 live points, which typically results in a couple thousand posterior samples for analysis, and allow the algorithm to run until reaching a tolerance criterion which is set to $1\times10^{-12}$. In the end, we retrieve the probability densities and the mixing ratio of each molecule after correction for their ionization cross section. We include 23 species in the database for the fitting. Since the signals above 47 amu are near the noise level, the base peaks of the 23 molecules are all below 47 amu. The selection of these 23 species is based on the current understanding of the $N_2$-$CH_4$ chemistry and the signals on the mass spectra. These species and their relative fragmentation patterns from the NIST are listed in Table S1 (Supporting information). One exception is methanimine ($CH_2NH$), which is detected in Titan's upper atmosphere by INMS[10,11] and identified in $N_2$-$CH_4$ plasma experiments.[15,16] However, there is no reliable fragmentation pattern for $CH_2NH$. We first fit our mass spectra using the Monte Carlo approach with the other 22 species, and then estimate the abundance of $CH_2NH$ by assuming that the underfitting for the peak at 29 amu (molecular ion peak for $CH_2NH$) is derived from $CH_2NH$. We make this assumption because 29 amu has no other reasonable molecules; the contribution of $C_2H_5^+$ is already accounted from the ethane fragmentation pattern, and $CHO^+$ is unlikely because of the low oxygen abundance in our experiments (the mixing ratios of $H_2O$ and $O_2$ are ~$2\times10^{-4}$ or ~$1\times10^{-5}$, respectively). Although the uncertainty of $CH_2NH$'s abundance may be larger than other species due to lack of reliable fragmentation pattern, the current approach can still provide a good estimation.

3. RESULTS AND DISCUSSION

*3.1. Mass spectral deconvolution results*

Figure 3 shows an example of the mass spectral deconvolution result using our Monte Carlo approach for the 10%-$CH_4$-10-sccm-2-Torr experiment. The upper panel (A) shows the subtracted mass spectrum (black outline bars) with the fitted spectrum contributed by each species from the model (colored segments); the lower panel (B) shows the mixing ratio of each molecule retrieved by the model. The black circles in Figure 4B represent the median mixing ratios, and the width of the shaded color region represents the probability density from the fitting simulations. In this study, we use 23 species (Table S1) to fit the mass spectrum from 1 to 46 amu. The retrieved relative abundances and their standard deviations (1σ) are listed in Table S2. Among these 23 species, 8 of them have mixing ratio higher than 100 ppm and the mixing ratios of the other 15 species are lower than 0.1 ppm. In addition, the uncertainties of the estimated mixing ratios for the 8 major species are all less than 2%, while that of the other 15 species vary from 0.1% to 90%. The higher uncertainties in some cases (for some of the minor species) are likely due to overlapping of the fragments from several species (e.g., $C_2H_2$, $C_2H_4$, $C_3H_4$, and $C_3H_6$ all have fragments at 26 amu).

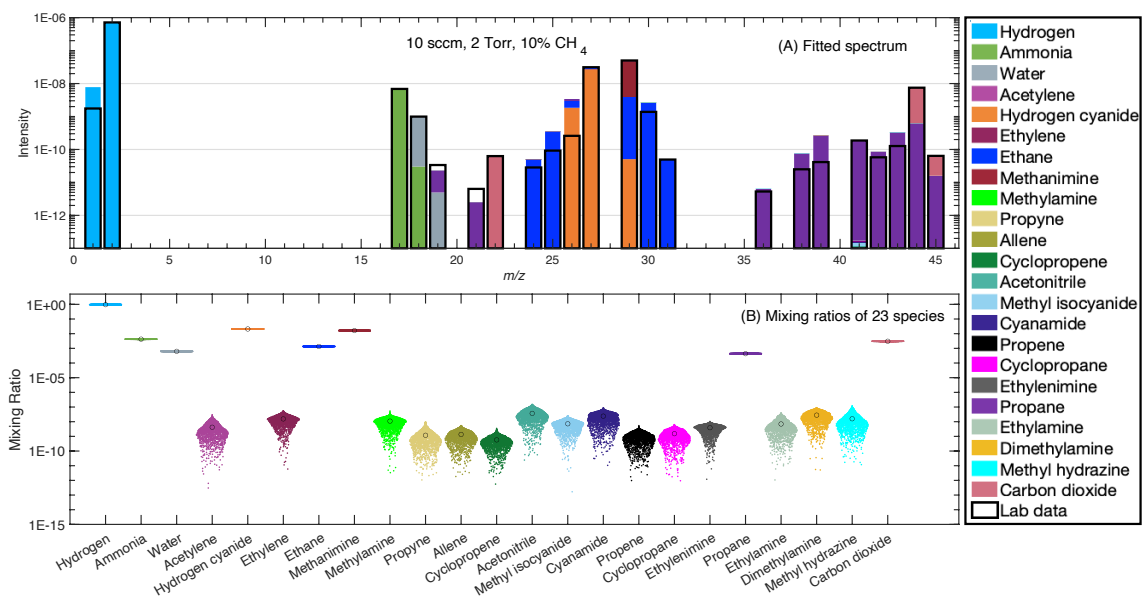

Figure 3. (A) Example of mass spectral deconvolution result using our Monte Carlo approach for the subtracted mass spectrum in the 10%-CH$_4$-10-sccm-2-Torr experiment. Black outline bars represent the subtracted mass spectrum from lab measurements. Colored segments represent the contribution of each species to the mass channel as calculated by the model. (B). The mixing ratio of each species retrieved by the model. The black circles represent the median mixing ratios, while the width of the shaded color region represents the probability density from the fitting simulations. The flat "lines" for the major species (>100 ppm) indicate that the fitting simulations give consistent mixing ratios for these major species.

Using the same approach, we fit the subtracted mass spectra for all our experiments. The mass spectral deconvolution results and the retrieved mixing ratios with their probability density are shown in Figures S1-S12 (Supporting Information). The retrieved relative abundances and their standard deviations (1σ) are listed in Table S2 (Supporting Information). The fitting results show that the same 7 molecules (H$_2$, NH$_3$, H$_2$O, HCN, CH$_2$NH, C$_2$H$_6$, and C$_3$H$_8$) always have higher mixing ratios and lower uncertainties in all our experiments (13 cases). Water could come from the buildup of water adsorption in the mass spectrometer chamber and/or from the accumulation of micro-leaks.[31] The other 6 species are the major gas products in our experiments, with hydrogen as the most dominant species (mixing ratio>60%), followed by three nitrogen-containing molecules (NH$_3$, HCN, and CH$_2$NH) and then two hydrocarbons (C$_2$H$_6$ and C$_3$H$_8$). Using similar deconvolution approach, Bourgalais et al.[31] identified gas products in their closed cell Titan simulation experiments with EUV source (89.2 nm). They found dimethyldiazene ($C_2H_6N_2$) is the main nitrogen-containing organic molecule followed by $CH_3CN$ and HCN that are also present in our experiments. They identified butane ($C_4H_{10}$), acetylene ($C_2H_2$), and propyne ($CH_3C_2H$) as major hydrocarbons, and detected several oxygenated species due to higher water vapor in their system.[31]

*3.2. $CH_4$ mixing ratio affects gas products*

The mixing ratio of $CH_4$ in the initial $CH_4/N_2$ gas mixture affects the chemistry induced by the plasma, therefore impacting the composition of gas and solid products. Based on the mass spectral deconvolution results, we plot the relative abundance of the gas products against $CH_4$ concentration in Figure 4. As shown in Figure 4A, the total yield of the gas products increases significantly with $CH_4$ concentration, dominated by the increase of $H_2$. The yields of all non-$H_2$ products, nitrogen-containing compounds, and pure hydrocarbons ($C_xH_y$) all increase moderately with $CH_4$ concentration. The yield of the non-$H_2$ products is mainly contributed by the nitrogen-containing compounds, whose yields are at least 10 times higher than that of the hydrocarbons in these five experiments with different $CH_4$ concentration. Figure 4B and 4C show the relative abundances of five major species in the experiments: three nitrogen-containing compounds ($NH_3$, HCN, and $CH_2NH$) and two hydrocarbons ($C_2H_6$ and $C_3H_8$). For the nitrogen-containing compounds, HCN has a higher yield than $NH_3$ and $CH_2NH$ in all 5 experiments. The yields of HCN and $NH_3$ drop slightly, but that of $CH_2NH$ increases significantly with $CH_4$ concentration. On the other hand, the yield of hydrocarbons gradually increases with $CH_4$ concentration, and two major species ($C_2H_6$ and $C_3H_8$) follow the similar trend. The increase of $CH_2NH$ may contribute to the formation of solid particles as previous studies have suggested that $CH_2NH$ is an important precursor for the solid formation.[29,40-42]

The initial gas mixture contains only $N_2$ and $CH_4$, and the chemistry induced by plasma generates larger and more complex organic molecules in the gas and solid products. In the system, the initial $CH_4$ provides a carbon source for any new organic products, and any new bond formed between carbon and carbon or carbon and nitrogen will free one hydrogen atom. The free hydrogen atom can form $H_2$ molecule through two possible pathways (H+H+M→$H_2$+M or $^3CH_2$+H→CH+$H_2$).[43] Therefore, the yield of $H_2$ serves as an indicator of the production rate of new organic molecules in the experiment. Figure 4 shows that the yield of $H_2$ increases more substantially with $CH_4$ concentration than that of non-$H_2$ gas products in our experiments. The difference in yields indicates that at higher $CH_4$ ratios, the organic molecules formed during the experiments end up mostly in the solid products rather than the gas phase.

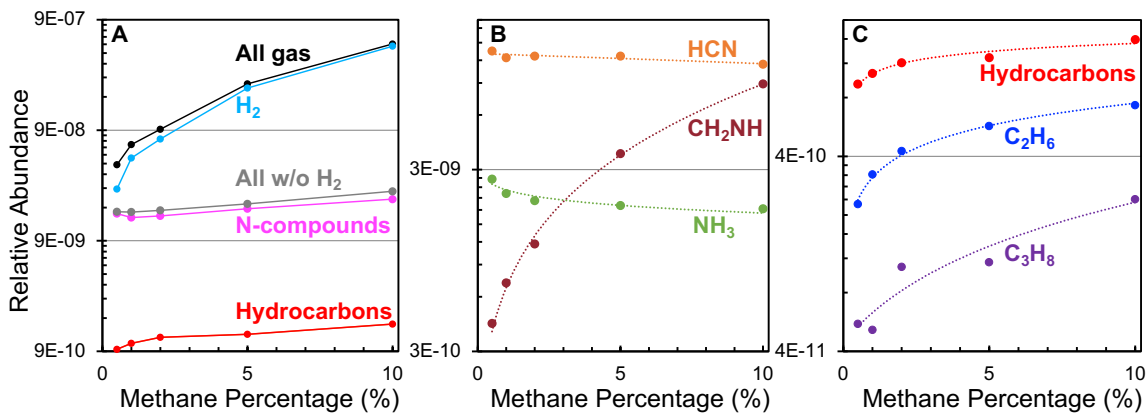

Figure 4. The relative abundance of the gas products changes with $CH_4$ concentration in the initial $CH_4/N_2$ gas mixture. "**All gas**" represents the total abundance of 23 gas products in each experiment, "**All w/o $H_2$**" represents the abundance of the other 22 gas products not including $H_2$, "**N-compounds**" represents that of 11 nitrogen-containing compounds, and "**Hydrocarbons**" represents that of pure hydrocarbons ($C_xH_y$). The uncertainties of the relative abundance for the six major gas products in the figure are less than 2%, whose size is smaller than the symbols.

*3.3. $CH_4$ mixing ratio affects production rate and elemental composition of the solids*

We run the experiments with the different $CH_4$ ratios for 72 hours and collect the produced solid samples after experiments. We weigh the solid material and calculate the haze production rate in these experiments as shown in Figure 5A. Since it is impossible to retrieve all the produced solid samples from the chamber wall, the production rates shown here likely reflect a lower limit. As $CH_4$ concentration increases from 0.5% to 10%, the haze production rate increases from 2.2 to 10.7 mg h$^{-1}$. Although it is not a strict linear relationship, the haze production rate increases gradually with the $CH_4$ mixing ratio, which is consistent with the substantial increase of $H_2$ production in the gas phase. In these experiments, the increase of $CH_4$ concentration promotes the production of both the gas and solid products. As $CH_4$ concentration increases from 0.5 to 10% (20 times increase), the non-$H_2$ gas products only increase 1.5 times while the solid products increase ~5 times, suggesting that most of the consumed $CH_4$ incorporates into the solid products.

Figure 5B shows the elemental compositions of the solid products in the experiments with different $CH_4$ concentrations. As $CH_4$ increases, the mass percentage of carbon in the solid products increases from 43.5% to 51.8%, while that of nitrogen decreases from 49% to 40%. There is a slight increase of hydrogen content (4.5 to 5.7%), and the oxygen content is less than 3% across all the samples. Oxygen and water contamination are well known problems for tholin production and analysis.[4] We employ Ultra-high vacuum system and dry-$N_2$ glove box in the process of sample production and collection to minimize such contamination. The intrinsic oxygen in the solid sample may only account for a small portion, and the oxygen contamination in the solid samples is most likely from the water adsorbed when the samples are weighed for the elemental analysis. The level of contamination (<3%) is relatively low compared to previous studies.[13,44]

The carbon to nitrogen (C/N) ratio in the initial gas mixture rises from 1:398 to 1:20 as $CH_4$ concentration increases from 0.5% to 10%, which leads to 5-fold growth of the solid production rate. One would expect significant growth of C/N ratio in the solid products. However, the elemental analysis shows that the C/N ratio only increases from 1:1 to 3:2, suggesting that the mass increase of the solid products is contributed by both carbon and nitrogen. In other words, additional $CH_4$ in the initial gas mixture promotes the incorporation of both carbon and nitrogen into the solid products. This is consistent with the composition of the gas products; there are >10 times more nitrogen-containing compounds than pure hydrocarbons ($C_xH_y$) produced in the gas products. Especially $CH_2NH$, whose yield increases substantially with $CH_4$ ratio while other major nitrogen-containing compounds (HCN and $NH_3$) decline slightly, is probably responsible for the enhancement of the solid production and the incorporation of both carbon and nitrogen (in 1:1 ratio).

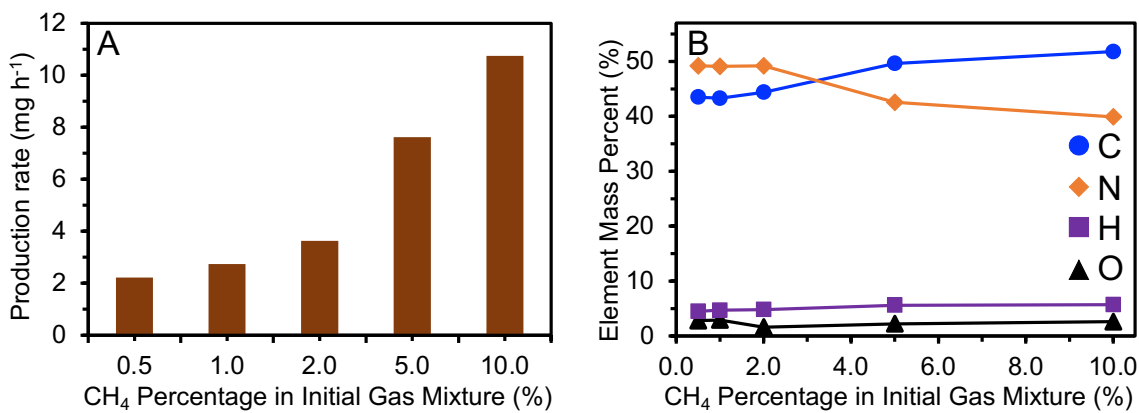

Fig. 5 The production rate (A) and the elemental compositions (B) of the solid products in the experiments with different $CH_4$ mixing ratios. The production rates likely represent a lower limit. The error bars are ~0.5% for elemental compositions from replicate runs and are the same size or smaller than the symbols.

*3.4. Gas flow rate and pressure affects gas products*

As the 10% $CH_4$ experiment has the highest gas and solid production rate among the five $CH_4$ concentrations analyzed here, we further investigate the effect of the gas flow rate and pressure on the gas products with 10% $CH_4$ in the initial gas mixture. Figures 6A-B show that the gas products change with flow rates at pressures of 3 and 5 Torr. At 3 and 5 Torr, the total yield of gas products decreases with flow rate, mainly because less $H_2$ is produced at higher flow rate. As we discussed in Section 3.2 and 3.3, the yield of $H_2$ in the experiments can indicate the amount of new organic products, the majority of which condense into the solid phase. Therefore, the decline of the $H_2$ production suggests that the production rate of the solid products probably decreases with flow rate at given pressure, which is likely caused by different exposure time of the reactant gas mixture to the plasma discharge and/or different residence time of the gas mixture in the reaction chamber. Hörst et al.[16] also observed that the total volume of aerosol produced decreases with increasing flow rate in their Titan simulation experiments.

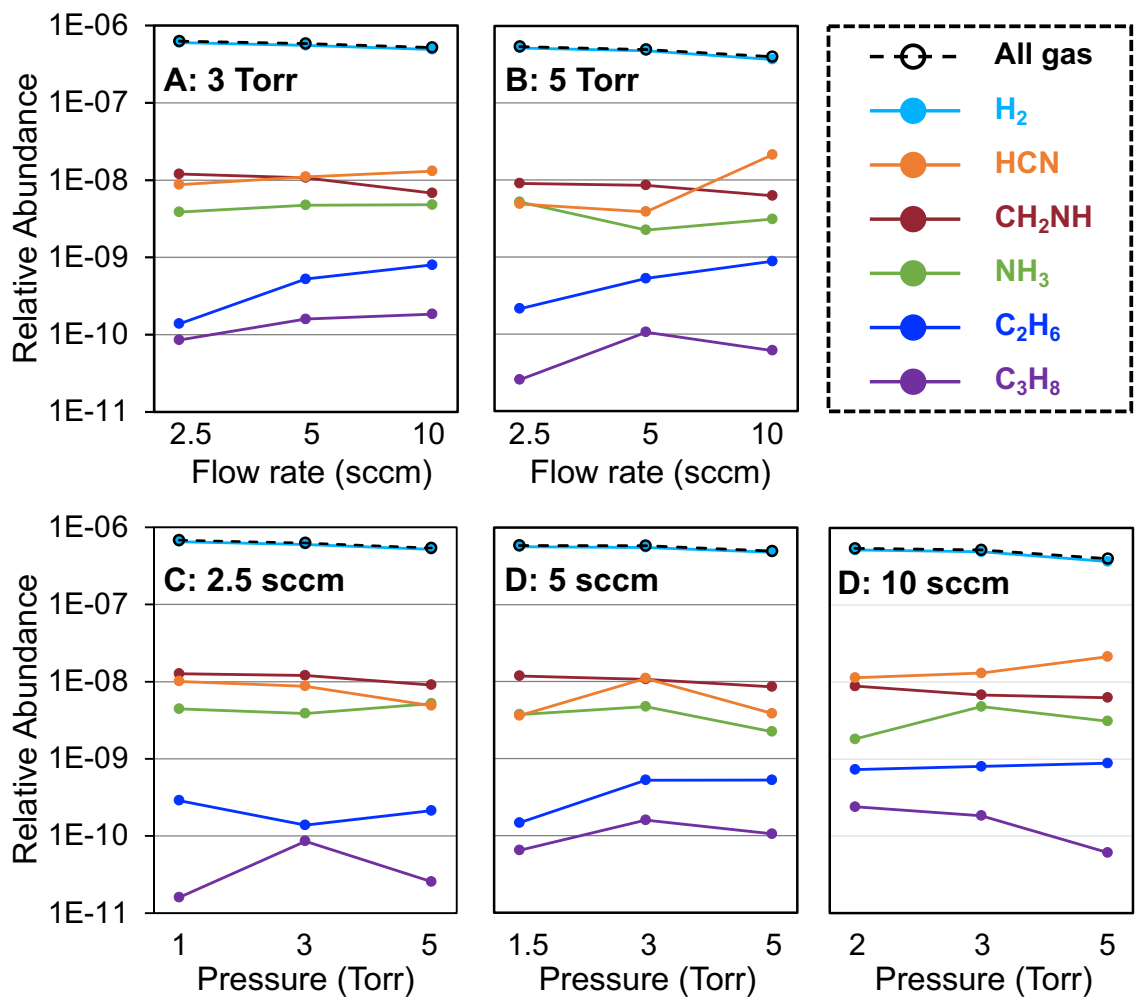

Figure 6. The yield change of the gas products with flow rates at fixed pressure of 3 Torr (A) and 5 Torr (B) in the 10% $CH_4$ experiments. The yield change of the gas products with pressure at fixed flow rates (C: 2.5, D: 5, and E: 10 sccm) in the 10% $CH_4$ experiments.

In our experiments, the exposure time of the reactant gas mixture to the plasma discharge can be calculated based on Equation 1;

$$t_1 = \frac{[2 \times \pi (R^2 - r^2) \times H] \times P_1}{Q \times P_0} \qquad \text{Eq.1}$$

Where $P_0$ is standard pressure (1 bar or 760 Torr); $P_1$ is the pressure in the chamber; R and r represent the radius of the plasma cylinder and the electrode, respectively; H is the height of the plasma cylinder; Q is the flow rate; and the factor of 2 is for two electrodes in our chamber. The plasma cylinder radius (r) decreases with the pressure ($P_1$) in the chamber; so, at given pressure, r keeps the same and the exposure time ($t_1$) decreases with flow rate (Q). The residence time of the gas mixture in the reaction chamber can be calculated following Equation 2;

$$t_2 = \frac{V \times P_1}{Q \times P_0} \qquad \text{Eq.2}$$

Where $P_0$, $P_1$, and Q are the same as Eq.1; and V is the volume of the reaction chamber. Based on Eq. 2, the gas residence time decreases with flow rate when the pressure ($P_1$) in the chamber remains the same. The average exposure time and the gas residence time are listed in Table 2. At given pressure, the gas mixture in the lower flow rate experiment is exposed to the plasma longer and also stays in the reaction chamber longer, which allows the reaction to proceed more thoroughly and leads to the formation of more organic products.

**Table 2.** The exposure time of the gas to the plasma discharge and the residence time of the gas in the reaction chamber.

| Flow rate (sccm) | 2.5 | | | 5 | | | 10 | | |
|---|---|---|---|---|---|---|---|---|---|
| Pressure (Torr) | 1 | 3 | 5 | 1.5 | 3 | 5 | 2 | 3 | 5 |
| Exposure time (s) | 5.2 | 4 | 2.4 | 2.5 | 2 | 1.2 | 1.3 | 1 | 0.6 |
| Residence time (s) | 64 | 192 | 320 | 48 | 96 | 160 | 32 | 48 | 80 |

Figures 6C-E show that at fixed flow rate (2.5, 5, and 10 sccm), the total yield of gas products decreases when the pressure in the chamber increases, also largely caused by the decline of the $H_2$ production. Similarly, the decline of the $H_2$ production likely indicates that the production rate of the solid products decreases with pressure at fixed flow rate. As shown in Table 2, at fixed flow rate, the higher-pressure experiments have longer residence time but short exposure time. However, the production rate of $H_2$ and solid products drops at higher pressure, indicating that the exposure time of the gas to the plasma discharge is more critical for the $N_2$-$CH_4$ chemistry to happen.

The yields of major nitrogen-containing products ($NH_3$, HCN, and $CH_2NH$) and hydrocarbons ($C_2H_6$ and $C_3H_8$) are also plotted in Figure 6. Figures 6A-E show that the yield of $CH_2NH$ decreases with flow rate or pressure, following the same trend as $H_2$ production in the experiments. This again suggests that $CH_2NH$ may play an important role in solid formation. On the contrary, the yield of $C_2H_6$ increases with flow rate or pressure in general with one exception shown in Figure 6C. There is no clear trend in the yields of the other three species across the experiments. Although each individual product shows different trend in different conditions (Figure 6), the total yields of the nitrogen-containing compounds are at least 10 times higher than those of the hydrocarbons. Among all nitrogen-containing compounds, HCN and $CH_2NH$ are the most abundant species, which have comparable yield in most of the experiments.

*3.5. Comparison with previous studies and implication for Titan*

Previous Titan atmosphere simulation experiments have investigated how the gas and solid phase chemistry is impacted by different experimental conditions, including $CH_4$ concentration, pressure, and flow rate.[13-16,18] Note that these simulation experiments usually use different experimental setups and analysis techniques. With the increase of $CH_4$ concentration, the yield of the gas products increases in our experiments. This trend is in agreement with previous plasma experiments, either with hot plasma[16] or with cold plasma.[14,15] $H_2$ is the dominant gas product in our experiments and the amount of $H_2$ serves as an indicator for organic products for our experiments. The production of $H_2$ increases with increasing initial $CH_4$ concentration (from 0.5 to 10%) in our experiments, while Carrasco et al.[15] have reported that $H_2$ first increases with $CH_4$ concentration but reaches an asymptote for initial $CH_4$ concentrations higher than 5%. We find that HCN is the most abundant non-$H_2$ product, and almost all previous experiments agree with this observation (see Figure 7 in Hörst et al. 2018[16]). Remarkably, the ratio of $H_2$/HCN in our 5% $CH_4$ experiment is nearly identical to that observed in Titan's atmosphere by Cassini INMS.[11,45]

Carrasco et al.[15] have reported an intermediate $CH_4$ ratio (5%), where the dominant gas composition transitions from nitrogen-containing species to hydrocarbons. Although we do see an increase of hydrocarbons with $CH_4$ mixing ratio in our gas products, the nitrogen-containing gas products always have much higher yield in our experiments with $CH_4$ from 0.5 to 10%, which is also observed by Hörst et al.[16] in their measurement of gas phase products.

Previous Titan atmosphere simulation experiments have demonstrated that the nitrogen-containing gas products are predominantly nitriles.[2,14,16,46] Consistently, we identify HCN as the most abundant non-$H_2$ product and also see acetonitrile ($CH_3CN$). Additionally, we observe that $NH_3$ and $CH_2NH$ are the other two major nitrogen-containing gas products, which are both detected in previous Titan experiments[15,47] and in Titan's atmosphere.[11,48] Besides HCN, these two species are also important precursors for prebiotic chemistry and for nitrogen chemistry in Titan's atmosphere.[40-42] The substantial yield increase of $CH_2NH$ with $CH_4$ ratio in our experiments may indicate its special role in incorporating both carbon and nitrogen into the solids and enhancing the efficiency of gas to particle conversion. $CH_2NH$ is known as a very reactive molecule and can easily polymerize or react with other species. Previous studies have demonstrated that it participates in many reactions during Titan's nitrogenous aerosol formation.[40-42] $CH_2NH$ is suggested as a "characteristic product" of plasma experiments and is important for the formation of heavy ions in Titan's ionosphere.[15] Recently, $CH_2NH$ is also detected in the experiments with EUV photons at 73.6 nm[29], demonstrating it might play a key role as an intermediate toward the formation of complex N-bearing organic molecules.

We find that the solid production rate increases when $CH_4$ ratio increases from 0.5 to 10%, while Sciamma-O'Brien et al.[13] reported that in their RF plasma experiments ($CH_4$ ratio: 1 to 10%), tholin production rate first increases with $CH_4$ concentration, reaches a maximum (at ~5% $CH_4$) and then decreases. They suggested that the increasing amount of atomic hydrogen in the plasma as well as the increase in aliphatic contributions in the tholins might inhibit tholin production in their experiments. Our tholin production rate is lower in general compared to their results. It is possible that our experiments have not reached the inhibiting stage with 10% $CH_4$. We will further investigate the production rate with higher $CH_4$ ratio in the future. In terms of the solid compositions, we observe that the H/C ratios remain the same (~1.3±0.1) regardless of $CH_4$ ratio. The independence of H/C ratio to $CH_4$ ratio is also observed by Hörst et al.[16] in their AMS measurement of the solid particles. The N/C ratios in our solid products are 0.97±0.01 when $CH_4$ ratio is 0.5, 1, and 2%, and decreases to 0.73 and 0.67 with 5% and 10% $CH_4$, respectively. Our N/C ratio is the highest compared to previous plasma simulations with the same $CH_4$ ratio (see Figure 17 in Hörst et al. 2018[16]). In all previous Titan atmosphere simulation experiments, the vast majority of them find N/C ratios between about 0.35 and 0.9. The 0.97±0.01 N/C ratio is highest regardless of $CH_4$ ratio and energy source, with only one exception ($CH_4$:$N_2$ 1:999 in McDonald et al. 1994[49]). The greater degree of nitrogen incorporation in the low temperature Titan simulation experiments suggests that the cold environment in Titan's atmosphere may promote nitrogen incorporation.

The energy density of the plasma energy source in our experiments[22-24] is about 170 W/m$^2$, which is about 10 times higher than the solar UV radiation reaching the top of Titan's atmosphere (14-17 W/m$^2$).[50] In addition, the energy inputs of charged particles (i.e., cosmic rays, magnetospheric protons, and electrons) in the upper atmosphere can exceed the solar UV when Titan moves across the Saturnian magnetopause and is fully exposed to solar wind.[4,51] For a given gas mixture (10% $CH_4$ in $N_2$ in our experiments), varying the pressure or flow rate affects the time of the gas exposed to the energy source. Longer exposure time leads to greater gas and solid production rate, which is consistent with previous study.[16] Although the pressure and/or flow rate have smaller effects on the gas and solid production rate compared to the initial gas composition, the result emphasizes the role of the energy flux in Titan atmospheric chemistry. Imanaka et al.[18] reported that pressure also affects the chemical structures in the solid products when the pressure varies dramatically (from 13 Pa to 2300 Pa). However, no significant difference of the gas products is observed in our experiments, probably due to the smaller pressure change among our experiments (1 to 5 torr).

4. CONCLUSIONS

In this study, we perform low temperature Titan atmosphere simulation experiments with $CH_4$-$N_2$ gas mixtures ranging from 0.5% to 10% $CH_4$ at different pressure and flow rate. We monitor the gas composition with in situ mass spectrometer and identify the gas products from deconvolution of the mass spectra with a Monte Carlo approach. The production rate and elemental composition of the solid products are also characterized. We find that the production rate of the gas and solid products increases with the $CH_4$ ratio. $H_2$ is the primary gas product and serves as an indicator of the production rate of new organic molecules in the experiment. In the non-$H_2$ gas products, the nitrogen-containing species have much higher yield than hydrocarbons; the N/C ratio of the solid products appears to be the highest compared to previous plasma simulations with the same $CH_4$ ratio, demonstrating the important role of nitrogen incorporation in Titan's atmospheric chemistry. The yield of $CH_2NH$ increases substantially with $CH_4$ ratio while that of HCN and $NH_3$ only change slightly. This indicates that $CH_2NH$ may be responsible for incorporating both carbon and nitrogen into the solid particles, increasing the conversion efficiency from gas to particle. The pressure and flow rate mainly affect the time of the gas mixture exposed to the energy source, therefore impacting the $N_2$-$CH_4$ chemistry initiated by the plasma discharge.

**Acknowledgements**

We gratefully acknowledge support from NASA grants 80NSSC20K0271 and 80NSSC19K0903.

## Supporting Information

Supporting Table S1-S2 and Figure S1-S12 is available free of charge at https://pubs.acs.org/doi/10.1021/acsearthspacechem.xxxxxxx.

**Table of Contents (TOC) Graphic**

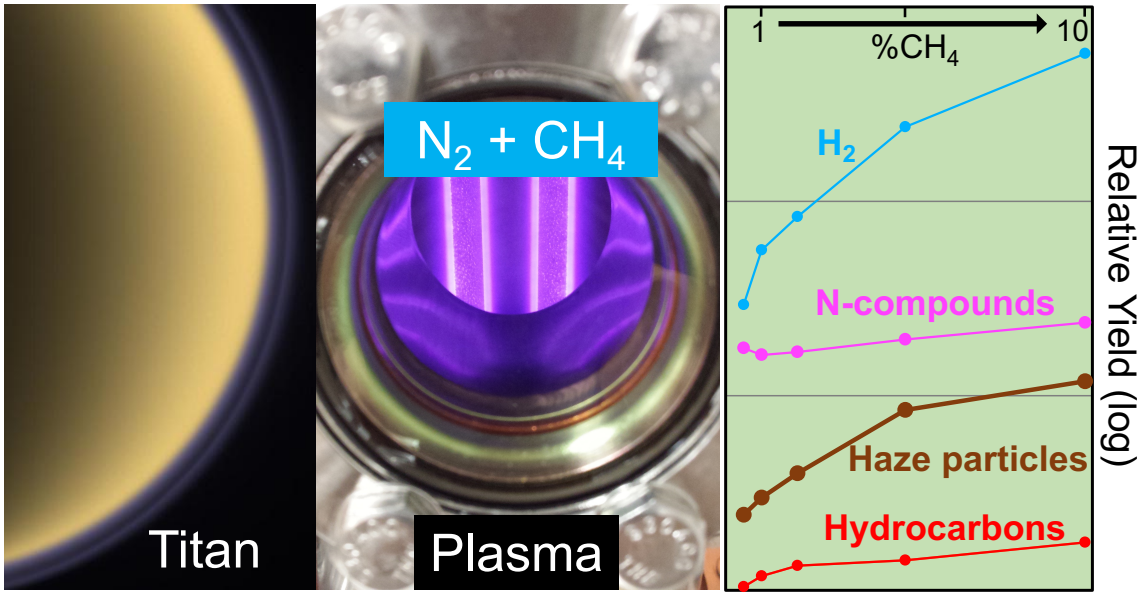